\documentstyle[12pt,dmtrieste,psfig]{article}

\def\mincir{\raise -2.truept\hbox{\rlap{\hbox{$\sim$}}\raise5.truept
\hbox{$<$}\ }}
\def\magcir{\raise -2.truept\hbox{\rlap{\hbox{$\sim$}}\raise5.truept
\hbox{$>$}\ }}
\def\rf{\par\noindent\hangindent 20pt}
\catcode`\@=11
\long\def\@makefntext#1{
\protect\noindent \hbox to 3.2pt {\hskip-.9pt
$^{{\ninerm\@thefnmark}}$\hfil}#1\hfill}                

\def\@makefnmark{\hbox to 0pt{$^{\@thefnmark}$\hss}}  

\def\ps@myheadings{\let\@mkboth\@gobbletwo
\def\@oddhead{\hbox{}
\rightmark\hfil\ninerm\thepage}
\def\@oddfoot{}\def\@evenhead{\ninerm\thepage\hfil
\leftmark\hbox{}}\def\@evenfoot{}
\def\sectionmark##1{}\def\subsectionmark##1{}}

\begin{document}

\centerline{\normalsize\bf NBODY/SPH simulations of individual
galaxies}\footnote{Proceedings of the Conference "DM-Italia-97", held in
Trieste(Italy), December 9-11,1997}

\vspace*{0.6cm}
\centerline{\footnotesize GIOVANNI CARRARO}
\baselineskip=13pt
\centerline{\footnotesize\it Department of Astronomy, Padova University}
\baselineskip=12pt
\centerline{\footnotesize\it }
\centerline{\footnotesize E-mail: carraro@pd.astro.it}
\vspace*{0.3cm}
\centerline{\footnotesize CESARIO LIA}
\baselineskip=13pt
\centerline{\footnotesize\it SISSA}
\baselineskip=12pt
\centerline{\footnotesize\it }
\centerline{\footnotesize E-mail: liac@sissa.it}
\vspace*{0.3cm}

\centerline{\footnotesize and}
\vspace*{0.3cm}
\centerline{\footnotesize FULVIO BUONOMO}
\baselineskip=13pt
\centerline{\footnotesize\it Department of Astronomy, Padova University}
\centerline{\footnotesize E-mail: buonomo@pd.astro.it}

\vspace*{0.9cm}
\abstracts
{We present preliminary results on galactic
Dark Matter (DM), halo structure, and galaxy evolution. We show how 
during the first Gyr of the evolution of a $10^{10} M_{\odot}$ dwarf
elliptica feed-back from stars (SN\ae~ and stellar winds) 
leads to  an extended constant density isothermal core with radius 
of 0.15 the virial radius $R_{200}$.
We also present first results on galaxy merging as a possibile
scenario to form ellipticals,
studying in particular how the details of the merging evolution vary as a 
function of the mass ratio of the interacting galaxies.}

\normalsize\baselineskip=15pt
\section{Introduction}
Numerical N-body and gasdynamical simulations have become a major tool
to investigate how galaxies formed and evolved. 
Massively parallel computers, like Cray T3D or T3E supercomputers, have
allowed to run high resolution simulations of large scale structure
formation (VIRGO or GIF projects), leading to impressive results
on the distribution of DM in different
cosmological scenarios (Jenkins et al. 1997). 

On galaxy scales, Navarro et al. (1996a) showed that galactic
DM halos should follow a universal density profile. In other words,
the violent, collisionless dynamical relaxation processes during the
formation of DM halos lead to equilibrium profiles with
similar shapes, independent of the halo mass, the initial density fluctuation
spectrum, and the adopted cosmological model.

However, recent studies (Persic et al. 1996; Burkert \& Silk 1997) have
found evidence, in real galaxies, of 
significant departures
from this universal profile both in the inner and in the outer galactic
regions.

\begin{figure*}[top]
\vbox{\vsize=10cm\vskip-2.1cm%
\centerline{\psfig{file=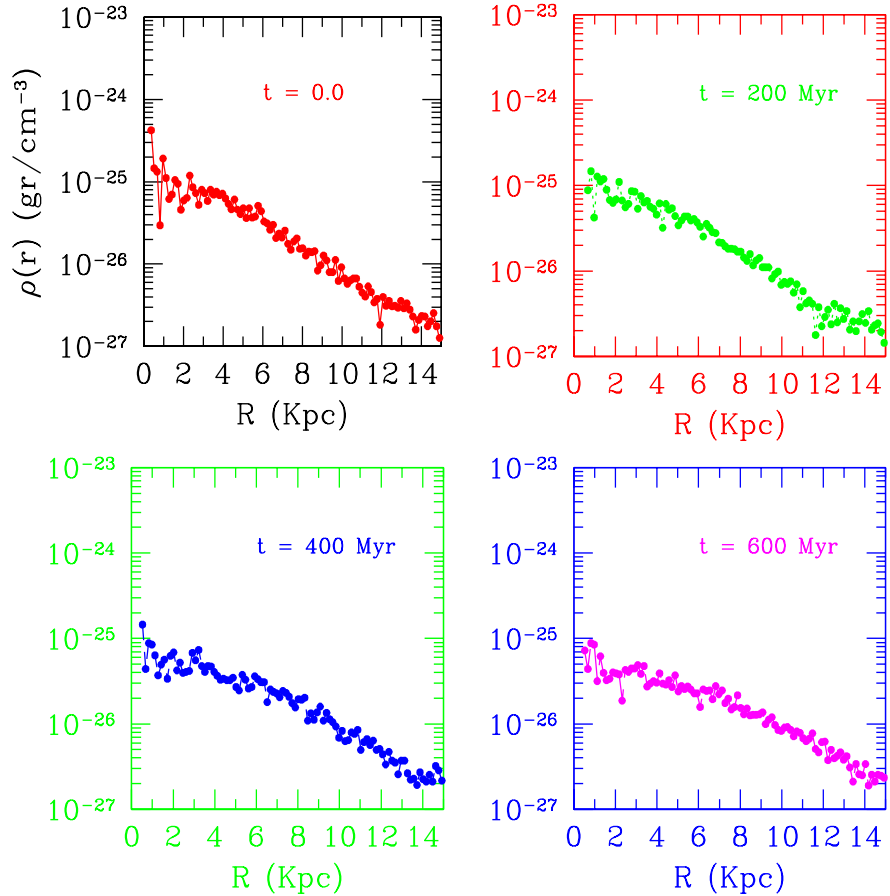,height=12cm,width=12cm}}
\caption{Time evolution of the DM density profile.}}
\end{figure*}

In particular, the presence of extended DM cores seems to be a firm
result.
The occurence and the development of cores with  costant
density is not yet completely understood. It has been suggested that
the discrepancy between observations and 
Cold Dark Matter (CDM) predictions could be solved by assuming
secular processes in the baryonic component which may also affect the
innermost halo regions. An analytic approach used
by  Navarro et al. (1996b) has suggested that such processes might produce 
core only in low mass galaxies.

In this contribution we present a fully Nbody
simulation
of the formation of a dwarf elliptical galaxy to investigate whether
baryonic feedback can  actually be the cause 
of the cores developments.

Tothis aim we use our own TreeSPH code (for details see Carraro et al 1997 
and  Lia et al 1998).

\section{DM haloes of Ellipticals}
Bertola et al (1993), using the gaseous component's rotation curve of the 
gas for a 
sample of giant ellipticals, showed that the halos surrounding  elliptical 
galaxies have characteristics similar to those of the spiral halos.
In particular the DM density profile in the inner region of the galaxies
is similarly much flatter than the baryons' profile, the two profiles 
intersecting at roughly the effective radius.

\begin{figure*}[top]
\vbox{\vsize=10cm\vskip-2.1cm%
\centerline{\psfig{file=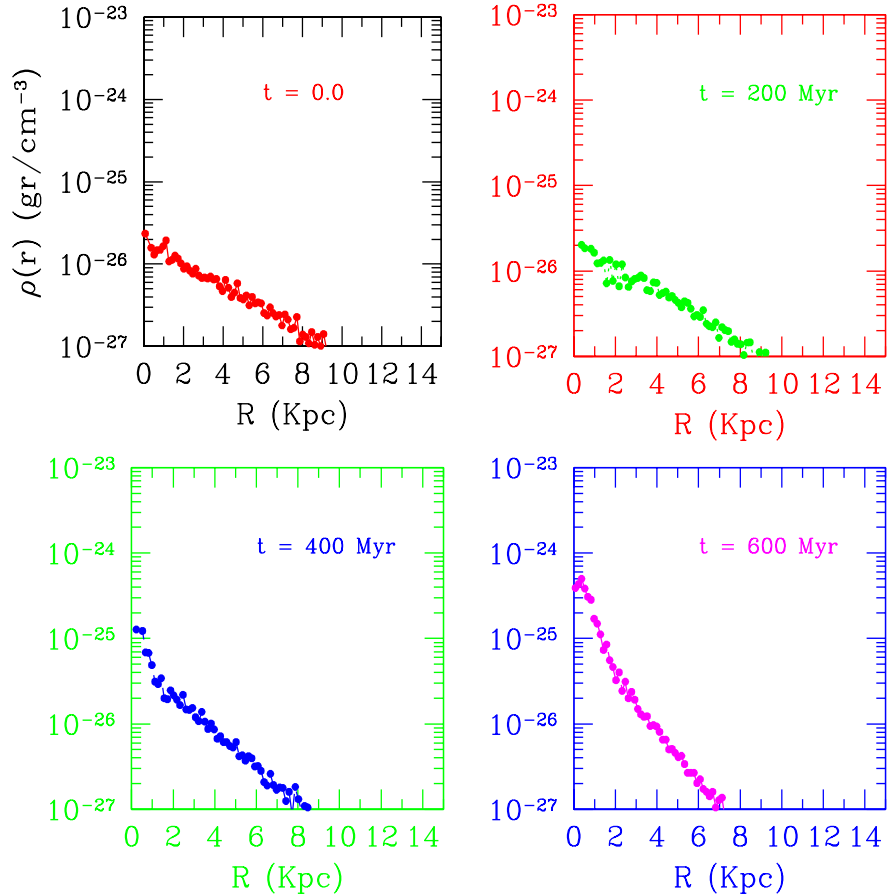,height=12cm,width=12cm}}
\caption{Time evolution of the baryons density profile.}}
\end{figure*}

In order to study the
formation of the core we have considered a  sperically symmetric 
virialized "primordial galaxy of "$10^{10}
M_{\odot}$ with a baryon
 fraction of $5\%$. The system is described by 8000
particle, half in form of baryons and half in form of DM.
Virialization is obtained by distributing dark particles according
to a $1/r$ density profile, with velocity components suitable to
produce the velocity dispersion necessary to support and stabilize
the system.
Gas is set on the top of dark particles with the same velocity
components.
At this point cooling switches on, and the gas falls in towards the center
forming stars. The time evolutions of the DM profile and of the baryon 
 profile are
 shown in Figs. 1 and 2, respectively; the situation at
the end of the run is presented in Fig. 3 (DM: open circles; baryons:
filled circles).

Looking at Figs 1-3 we clearly recognize that the DM density
profile gets flatter developing a core radius of about 2.0 kpc: this 
is similar to the effective radius as obtained by fitting the stars' 
distribution to 
a Hernquist profile. The baryons,  on the contrary, become more 
centrally concentrated, producing a steeper 
profile.

In order to investigate the origin of the core 
, we have
plotted in Fig. 4 the trend of the halo concentration,  defined as $\rho_{1}
/ \rho_{4}$, which is a
measure of the core development, against the SFR  at four epochs 
along the run.
There is a clear relation, suggesting that the dynamical feedback of 
star formation on the halo may be responsible
for the formation and development of the core . Work is in progress to see 
whether this scenario is appropriate also for more massive galaxies.

\begin{figure}
\centerline{\psfig{file=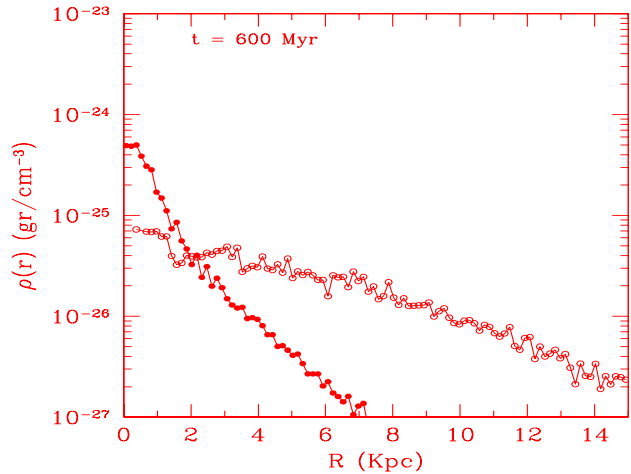,height=8cm,width=10cm}}
\caption{Baryons (filled circles) and DM (empty circles) density
profiles at the ned of the
simulation.}
\end{figure}

\begin{figure}
\centerline{\psfig{file=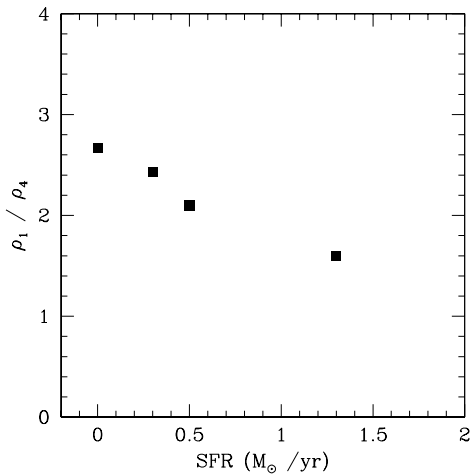,height=8cm,width=8cm}}
\caption{Core development as a function of the Star Formation rate.}
\end{figure}

\section{Galaxy Merging}
Many theoretical studies and numerical simulations have supported the
hypothesis
that the merging of two disk galaxies could be the 
dynamical process leading to the formation of an elliptical galaxy.
On the observational side, this hypothesys is supported by the presence, in many
ellipticals, of several photometric or kinematical peculiarities, like boxy
or disky departures of the isophotes from an  elliptical shape, 
counterrotating cores, the presence of shells, etc.

In the simulations described here, each galaxy is obtained 
from the adiabatic collapse of a rotating gas plus  sphere whith 
initial $r^{-1}$ density profile. Each galaxy 
consists of
1824 particles, equally distribuited between gas and DM.
We let the system evolve until virial equilibrium is reached.
Then we set the two galaxies along a specified orbit one around the
other.\\
As noted by Toomre \& Toomre (1972), the most interacting galaxies 
probably have highly eccentric orbits and are coming together for the first
time only now; so $0 \ll e \le 1$. 
On the other hand, a pair of galaxies in an
extended distribution of dark matter, could be accelerated to a nearly 
hyperbolic encounter ($e \ge 1$). So a good compromise between these two 
possibilities is to consider a parabolic orbit ($e=1$).\\ 
We assign the initial orbit in terms of the masses of the two galaxies,
of the eccentricity $e$, the aperture $p$ and the initial anomaly $\phi$. 
Then we calculate the initial velocities and initial positions of the 
particles in the center of mass frame. The parameters of the two encounters 
(in code units) are summarized in Table 1.

\begin{figure*}
\centerline{\psfig{file=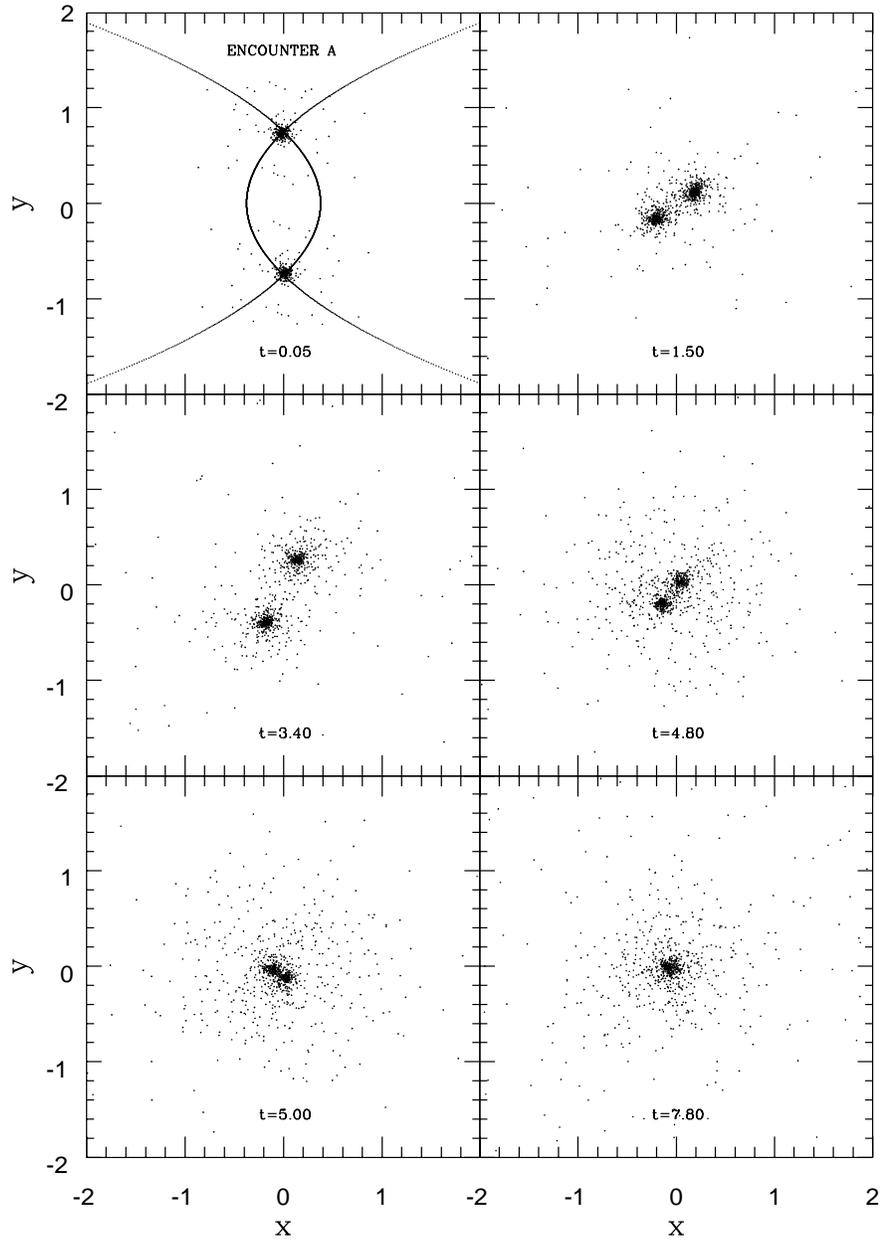,width=14cm,height=18cm}}
\caption{Merging evolution for the case A.}
\end{figure*}

\begin{figure*}
\centerline{\psfig{file=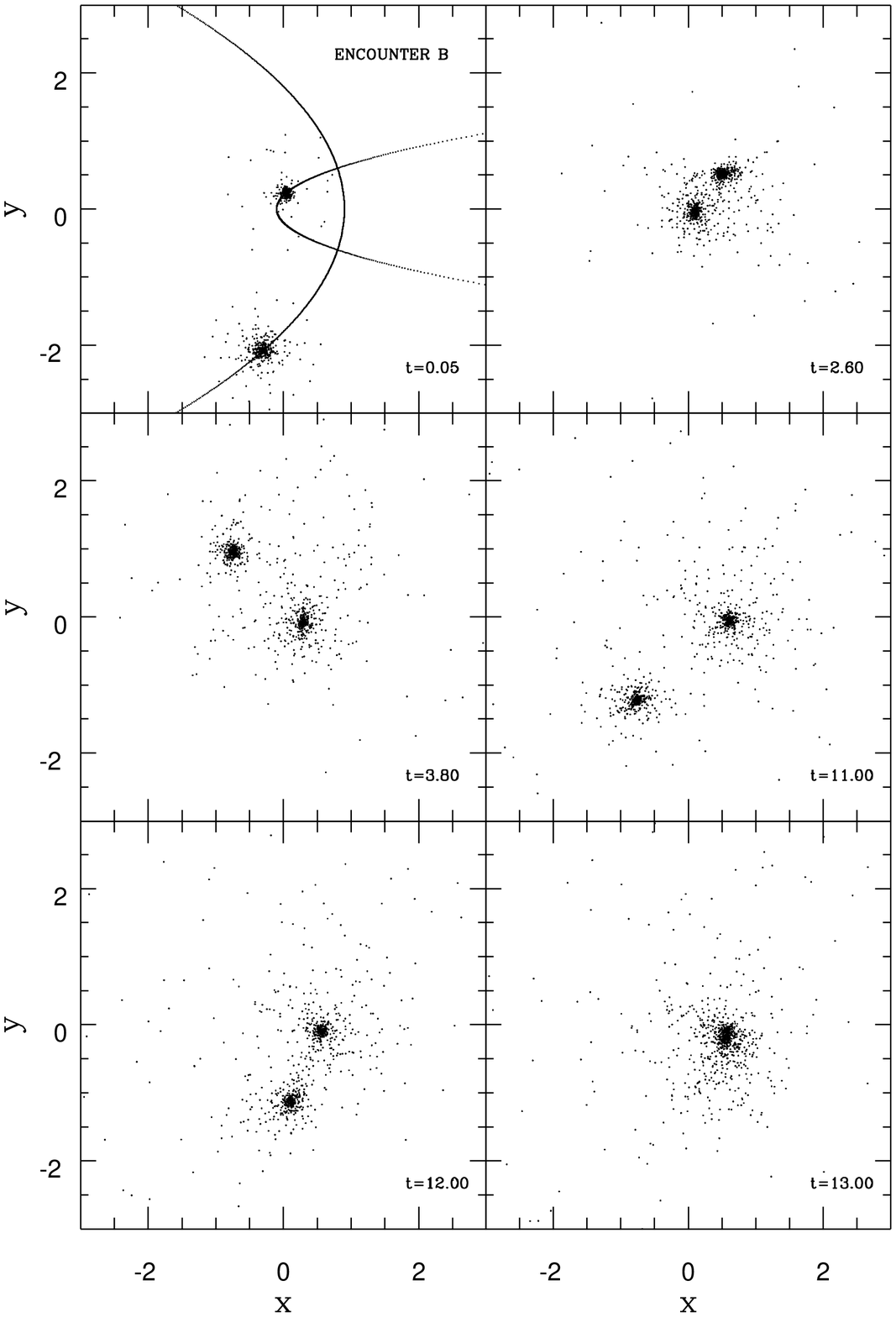,width=14cm,height=18cm}}
\caption{Merging evolution for the case B.}
\end{figure*}

\begin{table}
\begin{center}
\caption{Parameters for the merging simulations.}
\begin{tabular}{|c|c|c|c|c|c|} \hline
&&&&& \\
Model & $M_{1}$ & $M_{2}$ & $\phi$ & $\epsilon$ &
$p$ \\  &&&&& \\ \hline 
 & & & & & \\

A & 0.5 & 0.5 & $- \pi/2$ & 1.5 & 1.0   \\
&&&&& \\ \hline
&&&&& \\
B & 0.9 & 0.1 & $- 11/20 \pi$ & 1.0 & 2.0 \\
&&&&& \\ \hline 
\end{tabular}
\end{center}
\end{table}

In the adopted coordinate system, the orbital plane 
coincides with the XY plane and the origin is in the center of mass of the two 
galaxies. The galaxies move about each other in a counterclockwise direction 
and the encounters are direct.\\
In encounter A (Fig.~5) the two galaxies have the same mass (M1=M2=0.5, in
code units),
The two galaxies approach for the first time at $t=1.5$ but then 
go far apart, departing from the initial keplerian orbit. After 
the first approach,  the {\em gas} of each galaxy exhibits the effects 
of tidal 
stripping in the form of moderate tails opposit to the sense of motion.
Around $t=4.5$ the two galaxies approach for the second time, starting a 
slow merging that eventually leaves a round-shaped remnant at around $t=7.8$ 

Encounter B (Fig.~6) is slower than encounter A. Galaxy 1 (mass 0.9)
remains quite 
stationary, while galaxy 2, with a lower mass, orbits around galaxy 1.
There is a close encounter around $t=2.6$, then the two galaxies
go far apart. After the first encounter, galaxy 2 continues its
motion around galaxy 1 until  it merges at around  $t=13.0$.
It is interesting to notice that the gas of
galaxy 2 is
more spread out  during the encounter B than during the  encounter A.

\section{Conclusions}
This paper is a report on two ongoing projects in   
Galaxy
Formation. Within the context of the monolithic collapse scenario we have
proposed a possible explanation for the formation of a DM core  in
a
dwarf elliptical, emphasising  the role of secular
processes related  to stellar feedback.
In the context of the merging scenario for the formation of ellipticals, we 
have presented 
preliminary results on the merging of two spiral-like galaxies for two 
cases with different mass ratios.

\section{Acknowledgements}

We thank Paolo Salucci, Massimo Persic, Luigi Danese and
Ezio Pignatelli for enlightening discussions and continuous support.

\bigskip 
\bigskip

\section{References}
\bigskip
\rf{Bertola F., Pizzella A., Persic M., Salucci P., 1993,
	{\it ApJ} 416, L45}

\rf{Burkert A., Silk J., 1997,
        {\it ApJL} in press ({\tt astro-ph/9707343})}

\rf{Carraro G., Lia C., Chiosi C., 1997,
        {\it MNRAS} in press ({\tt astro-ph/9712307})}

\rf{Jenkins A. et al, 1997,
	{\it ApJ} submitted ({\tt astro-ph/9709010})}

\rf{Lia C., Carraro G., Chiosi C., 1998
        {\it MNRAS} submitted}

\rf{Navarro J.F., Frenk C.S., White S.D.M., 1996a,
	{\it ApJ} 462, 563}

\rf{Navarro J.F., Eke V.R., Frenk C.S., 1996b,  
	{\it MNRAS} 283, L72}

\rf{Persic M., Salucci P., Stel F., 1996,
        {\it MNRAS} 281, 27}

\rf{Toomre A. \& Toomre J., 1972,
        {\it ApJ} 179, 623}

\end{document}